\documentclass[aps,pra ,a4paper,reprint,twocolumn,floatfix,superscriptaddress,notitlepage,usenames,dvipsnames,svgnames,table,nofootinbib,longbibliography]{revtex4-1}
\pdfoutput=1
\usepackage[utf8]{inputenc}
\usepackage[T1]{fontenc}
\usepackage{graphicx}
\usepackage{grffile}
\usepackage{wrapfig}
\usepackage{rotating}
\usepackage[normalem]{ulem}
\usepackage{amsmath}
\usepackage{textcomp}
\usepackage{amssymb}
\usepackage{capt-of}
\usepackage{verbatim}
\usepackage[autostyle,english=american]{csquotes}
\MakeOuterQuote{"}
\usepackage{parskip}
\usepackage{mathrsfs}
\usepackage[margin= 0.75in]{geometry}
\usepackage[braket, qm]{qcircuit}

\usepackage[english]{babel}
\usepackage{bbm}
\usepackage{etoolbox}
\usepackage{tcolorbox} 
\usepackage{fleqn} 
\usepackage{tikz}
\usepackage{amsthm}
\usepackage{amssymb}
\usepackage{mathtools}
\usepackage{bm}
\usepackage{dsfont}
\usepackage[font=small,labelfont=bf,justification=justified,format=plain]{caption}
\usepackage{subcaption}
\usepackage{thmtools}
\usepackage{thm-restate}
\usepackage{hyperref}
\hypersetup{
 pdfauthor={Brian Barch},
 pdftitle={NH CCs},
 pdflang={English},colorlinks=true,citecolor=teal,urlcolor=Periwinkle} 
\usepackage[capitalise]{cleveref}

\graphicspath{{images/}}

\begin{document}

\title{Locality, Correlations, Information, and non-Hermitian Quantum Systems}
\author{Brian Barch}
\email[email: ]{barch@usc.edu}
\affiliation{Department of Physics and Astronomy, and Center for Quantum Information Science and Technology, University of Southern California, Los Angeles, California 90089-0484, USA}
\date{\today}

\iftrue
\begin{abstract}
Local non-Hermitian (NH) quantum systems generically exhibit breakdown of Lieb-Robinson (LR) bounds, motivating study of whether new locality measures might shed light not seen by existing measures. In this paper we discuss extensions of the connected correlation function (CC) as measures of locality and information spreading in both Hermitian and NH systems. We find that in Hermitian systems, $\delta\rho = \rho-\rho_A\otimes\rho_B$ can be written as a linear combination of CCs, allowing placement of an LR bound on $\Vert\delta\rho\Vert_2$, which we show generically extends to an LR bound on mutual information. Additionally, we extend the CC to NH systems in a form that recovers locality, and use the metric formalism to derive a modified CC which recovers not just locality but even LR bounds in local $PT$-Symmetric systems. We find that even with these CCs, the bound on $\Vert\delta\rho\Vert_2$ breaks down in certain NH cases, which can be used to place a necessary condition on which local NH Hamiltonians are capable of nonlocal entanglement generation. Numerical simulations are provided by means of exact diagonalization for the NH Transverse-Field Ising Model, demonstrating both breakdown and recovery of LR bounds.
\end{abstract}
\fi

\maketitle

\section{Introduction}
Though there exist a number of approaches for studying conditional time evolutions, description as a non-Hermitian (NH) Hamiltonian \cite{ashida_non-hermitian_2020} reveals unique phenomena such as phase transitions without gap closing \cite{matsumoto_continuous_2020}, exceptional points \cite{matsumoto_embedding_2022, gopalakrishnan_entanglement_2021}, and Lieb-Robinson (LR) bound violations \cite{matsumoto_continuous_2020, ashida_2018, dora_quantum_2020, turkeshi2023}. LR bounds limit the speed of operator growth under non-relativistic local Hamiltonians \cite{liebFiniteGroupVelocity, Nachtergaele_2006, bravyi_lieb-robinson_2006} and local Lindbladians \cite{poulin_lieb-robinson_2010}, and are intimately related to many other results such as exponential decay of correlations \cite{hastingsSpectralGapExponential2006,Nachtergaele_2006} and generation of topological order \cite{bravyi_lieb-robinson_2006}, making their violation in local NH systems surprising. 

In contrast, evolution under NH Hamiltonians may generate less entanglement than under their Hermitian counterparts \cite{barch2023, gopalakrishnan_entanglement_2021, turkeshi2023}, indicative of limited information spread. This can be understood by interpretation of NH systems as arising from continuous weak measurement with postselection. While continuous local measurement is shown to preserve LR bounds \cite{poulin_lieb-robinson_2010}, postselection can reduce total system entropy but does not itself transmit information, bringing into question whether local NH systems truly transmit information nonlocally. Additionally, a subset of NH Hamiltonians, known as pseudo-Hermitian or $PT$-Symmetric, can be mapped to Hermitian Hamiltonians in a modified Hilbert space, where they generate Unitary evolution \cite{mostafazadeh_pseudo-hermiticity_2002, mostafazadeh_2003, ju_non-hermitian_2019, karuvade_2022}. When this mapping is locality-preserving, one would then expect similar locality properties and dynamics to exist in the original Hilbert space. These reasons motivates questioning whether local NH systems truly have long-range interactions, or if metrics designed for CPTP time evolution simply fail to capture the structure of locality in NH systems. We provide evidence towards the latter, by constructing a modified connected correlation function (CC) which obeys LR bounds even in local $PT$-Symmetric systems while still acting as a faithful measure of entanglement.

CCs are commonly used as locality measures, and are connected to a range of topics such as chaoticity \cite{gharibyan_characterization_2020} and criticality \cite{matsumoto_embedding_2022, turkeshi2023}. Previous work has shown an LR bound on standard \cite{bravyi_lieb-robinson_2006} and $n$-partite \cite{Tran_2017} CCs, and demonstrated a connection between CCs and entanglement measures such as mutual information \cite{Tran_2017, wolfAreaLawsQuantum2008, flam2023, scalet_computable_2021}. As we show here, the difference $\delta\rho=\rho-\rho_A\otimes\rho_B$ can be not only bounded but written as a sum of CCs, extending the previously mentioned CC LR bound to functions of $\delta\rho$. In particular, we show that mutual information can generically be upper bounded by a multiple of $\Vert\delta\rho\Vert_2$, and thus by an LR bound (Eq.~\ref{eq:info-LR-bound}). 

As the traditional CC loses its usual locality properties in NH systems, we instead consider two extensions. The first is a connected version of the correlator in Ref.~\cite{Sergi_2015}, and reduces to the standard CC in the Hermitian limit. The second is motivated by the metric formalism of $PT$-Symmetric systems \cite{mostafazadeh_pseudo-hermiticity_2002, mostafazadeh_2003, ju_non-hermitian_2019, karuvade_2022}, and is the CC in a modified Hilbert space in which the system is Hermitian. While both recover the desired locality properties and can be used as measures of entanglement, the latter is further shown to obey LR bounds in $PT$-Symmetric systems. Previous works have addressed modified disconnected correlators in NH systems, but these do not have the same locality properties as their connected counterparts \cite{Sergi_2015,matsumoto_embedding_2022, dora_quantum_2020}. To our knowledge this is the first extension of CCs to NH systems, and the first proof of any form of LR bound in local $PT$-Symmetric systems.

A background on relevant work in LR bounds, CCs, and NH systems is provided in Sec.~\ref{sec:BG}. In Sec.~\ref{sec:CCs} we construct the two modified CCs and discuss their properties. In Sec.~\ref{sec:CC-entang} we relate them to entanglement by writing $\delta\rho$ as a sum of CCs, and show that this term generically upper bounds mutual information. Proof of LR bounds in $PT$-Symmetric systems for the new CCs is given in Sec.~\ref{sec:LR}. In Sec.~\ref{sec:applications} we discuss two applications: bounding the class of $PT$-Symmetric Hamiltonians capable of generating long range entanglement, and measurement of CCs as POVMs. Finally in Sec.~\ref{sec:conclusion} we discuss significance of results and prospects for future study. The appendix contains various results including extension of the CC LR bound to the unequal time case (\ref{app:CCLR-uneq}), $n$-partite extension of modified CCs (\ref{app:n-part}), bounding mutual information in terms of $\Vert\delta\rho\Vert_2$ and decomposition of $\delta\rho$ in terms of modified CCs (\ref{app:entang-bound}), additional results on LR bounds (\ref{app:UinvLR}), and an example case of using a local $PT$-Symmetric Hamiltonian to generate long range entanglement (\ref{app:app-entang}). 

\section{Background}
\label{sec:BG}
\subsection{Lieb-Robinson Bounds}
\label{sec:BG-LR}
Even in nonrelativistic local quantum systems, there exists an effective lightcone limiting the spread of operator growth, known as the Lieb-Robinson (LR) bound ~\cite{liebFiniteGroupVelocity, Nachtergaele_2006, bravyi_lieb-robinson_2006, poulin_lieb-robinson_2010}. This bound states that for a quantum system evolving under a local Hamiltonian $H$,
\begin{align}
    \left\Vert [O_A(t),O_B ]\right\Vert \leq c N_\text{min} \Vert O_A \Vert \Vert O_B \Vert e^{-\frac{L-vt}{\xi}}
\end{align}

where $O_A(t) = U^\dag_t O_A U_t$, $U_t = e^{-i H t}$, and $O_A,O_B$ are any local operators initially supported on subsystems $A,B$ of distance $L$ apart. Additionally, $N_\text{min} = \text{min}\{\vert A\vert,\vert B\vert\}$, and $c,\xi$, and $v$ are system dependant constants.

This arises because, by definition, we can decompose a local Hamiltonian into a sum of local terms as $H = \sum_R H_R$ where $\{R\}$ are a set of subsystems of the quantum system whose locality is independent of system size. At time zero, the infinitesimal Heisenberg time evolution of $O_A$ is

\begin{align*}
    \dot O_A\big\vert_{t=0} = i [H, O_A] = i\sum_{R\cap A \neq \varnothing} [H_R,O_A]
\end{align*}

as commutator terms for which $R\cap A = \varnothing$ vanish. After some small time $dt$, as each $H_R$ is local, $O_A(dt)$ will, to first order, be local on subsystem $\cup_{R\cap A\neq\varnothing}R$, and the process repeats. 

This result was extended to show that one can also place an LR bound on the norm difference between an operator and its restriction onto a lightcone \cite{bravyi_lieb-robinson_2006}. For some initially local operator $O_A$, pick some $l$ and let $S$ be the "spacelike" region, i.e. the region of subsystems with distance at least $l$ from $A$. Then we can define the restriction of $O_A(t)$ onto a lightcone of radius $l$ around $A$ as
\begin{align}
    O_A^l(t) \equiv \text{Tr}_S [O_A(t)]\otimes\frac{\mathbb{I}_S}{\text{Tr}[\mathbb{I}_S]}
\end{align}

It was shown in Ref.~\cite{bravyi_lieb-robinson_2006} that for $\Vert O_A\Vert\leq 1$,

\begin{align}
\label{eq:op-lightcone}
\Vert O_A(t) - O_A^l (t) \Vert \leq c \vert A \vert e^{-\frac{l-vt}{\xi}}
\end{align}

for the same $c, v, \xi$ as the original LR bound.

Interestingly, the LR bound breaks down when $H$ is instead a local NH Hamiltonian~\cite{matsumoto_continuous_2020}. We repeat the argument here for completeness. Consider a Hamiltonian $H = \sum_R H_R + i\Gamma_R$ composed of Hermitian $H_R$ and $\Gamma_R$, acting on local regions $R$. The infinitesimal time evolution of $O_A$ at time zero under this Hamiltonian is
\begin{align*}
    \dot O_A\big\vert_{t=0} &= i H^\dag O_A -i O_A H\\
    &= i\sum_{R\cap A \neq \varnothing} [H_R,O_A]-\sum_R \{\Gamma_R, O_A\}.
\end{align*}

While $[H_R,O_A]=0$ for $R\cap A = \varnothing$ at $t=0$, the same does not hold for $\{\Gamma_R,O_A\}$, which can in general cause nonlocal growth of $O_A$, and breakdown of the LR bound. This breakdown is related to non-unitality of the evolution, as highlighted by considering product non-Unitary time evolution operator $U = U_A\otimes U_B$. Here,
\begin{align*}
    (U_A\otimes U_B)^\dag O_A (U_A\otimes U_B) = (U_A^\dag O_A U_A)\otimes (U^\dag_B U_B)
\end{align*}

which is no longer local on $A$.

\subsection{Correlation Functions}
\label{sec:BG-CC}
The traditional CC for arbitrary operators $O_1,O_2$ is given by 
\begin{align}
\label{eq:trad-cc-equal}
    \langle O_1,O_2\rangle_c = \langle O_1 O_2\rangle-\langle O_1\rangle\langle O_2\rangle
\end{align}

 where $\langle\ \cdot\ \rangle = \text{Tr}[\rho\ \cdot\ ]$ denotes expectation value w.r.t. some state $\rho$, which when ambiguous will be denoted explicitly by $\langle\ \cdot\ \rangle_\rho$. This can be extended to the unequal-time CC:
\begin{align}
\label{eq:cc-herm}
    \langle O_1(t),O_2(t')\rangle_c = \langle O_1(t) O_2(t')\rangle-\langle O_1(t)\rangle\langle O_2(t')\rangle
\end{align}
When $t=t'$ this is equivalent to Eq.~\ref{eq:trad-cc-equal} on time evolved state $\rho(t)$. 

\subsubsection{Connected Correlators and Mutual Information}
\label{sec:BG-CC-entang}

Unlike the disconnected correlator $\langle O_A O_B\rangle$, the CC is always zero for product states $\rho_A\otimes\rho_B$. In fact, the CC between any disjoint operators $O_A, O_B$ acting on subsystems $A,B$ can be upper bounded in terms of the mutual information between the two subsystems \cite{wolfAreaLawsQuantum2008}:
\begin{align}
\label{eq:info-geq-CC}
    \frac{\vert\langle O_A,O_B\rangle_c\vert^2}{2 \Vert O_A\Vert^2 \Vert O_B\Vert^2} \leq I(A;B)
\end{align}
This result has been extended to a family of $\alpha$-R\'enyi mutual information measures \cite{scalet_computable_2021, flam2023}. Additionally, it was shown that the relationship is bidirectional for $\alpha=1/2$, in that the R\'enyi mutual information can be upper bounded by a sum of CCs \cite{flam2023}. In Sec.~\ref{sec:CC-entang} we extend this result to traditional mutual information, and show that $I(A;B)$ itself can be upper bounded in terms of a sum of CCs in certain generic cases. This is achieved by bounding it in terms of $\delta\rho=\rho-\rho_A\otimes\rho_B$, itself a quantifier of entanglement, which be written as a sum of CCs. The resulting proof is an extension of a result from Ref.~\cite{Tran_2017} which shows that a state is product across a bipartition iff all CCs for operators on separate sides of the bipartition are zero. I.e. that
\begin{align}
\begin{split}
    \rho = \rho_A \otimes \rho_B &\Leftrightarrow \langle O_A,O_B\rangle_c=0 \\
    \forall O_A \in\mathscr{L(H}_A)&, O_B\in\mathscr{L(H}_B) 
\end{split}
\end{align}
\subsubsection{LR Bounds for Connected Correlators}
When $\rho$ exhibits finite correlation length, i.e. $\langle O_A,O_B\rangle_c \leq \tilde c\ e^{-\frac{L}{\chi}}$ for system-dependent correlation length $\chi$, the LR bound on operators can be extended to one on CCs \cite{bravyi_lieb-robinson_2006, Tran_2017}. This is the case in, e.g., ground states of gapped local Hamiltonians \cite{Nachtergaele_2006}. Using Eq.~\ref{eq:op-lightcone} we can bound
\begin{align}
\label{eq:CC-LR-bound}
\begin{split}
    &\big\vert \langle O_A(t),O_B(t)\rangle_c \big\vert\\
    &\leq \left\vert \langle O_A^l(t), O_B^l(t)\rangle_c \right\vert + c\left(\vert A\vert+\vert B\vert\right) e^{\frac{l-vt}{\xi}}\\
    &\leq \tilde c\ e^{-\frac{L-2l}{\chi}} + c(\vert A\vert+\vert B\vert) e^{-\frac{l-vt}{\xi}}\\
    &\leq \bar c\ e^{-\frac{L-2vt}{\chi'}}
    \end{split}
\end{align}
for $\bar c = \tilde c+c(\vert A\vert+\vert B\vert)$, $\chi'=\chi+2\xi$, and the optimal $l = (\chi vt+\xi L)/\chi'$. This extends to the $n$-partite case as well \cite{Tran_2017}. Additionally, as shown in Appendix~\ref{app:CCLR-uneq}, this generalizes to the unequal-time case as
\begin{align}
\label{eq:CC-uneq-LR-bound}
    \vert\langle O_A(t),O_B(t')\rangle_c\vert \leq \bar c\ e^{-\frac{L-v(t+t')}{\chi'}}
\end{align}

\subsection{Non-Hermitian Quantum Mechanics}

The paradigmatic conditional evolution motivating NH quantum mechanics is the no-jump trajectory \cite{ashida_non-hermitian_2020,brunsimplemodelquantum2002}. Consider a state $\rho$ evolving under the standard Lindblad equation:
\begin{align}
    \dot \rho = -i[H,\rho]+\sum_a \left( L_a \rho L_a^\dag - \frac{1}{2}\{L_a^\dag L_a, \rho\} \right).
\end{align}
While the first and third terms generate time evolution within a single quantum trajectory, the second term $L_a \rho L_a^\dag$ corresponds to jumps between quantum trajectories. Postselecting out such jumps effectively removes this term, and the resulting evolution can be described by an NH effective Hamiltonian 
\begin{align}
    \label{eq:H_eff}
    H_{\mathrm{eff}} = H -\frac{i}{2} \sum_a L_a^\dag L_a.
\end{align}
The resulting time evolution operator $U_t = e^{-i H_{\mathrm{eff}} t}$ is no longer Unitary, and the resulting time evolution is completely positive but no longer trace-preserving. Thus in order to interpret the output as states, the time evolution must be manually trace-normalized:
\begin{align}
\label{eq:mixed-evolution}
    \rho(t) = \frac{U_t \rho U_t^\dag}{\text{Tr}[U_t \rho U_t^\dag]}
\end{align}
This is equivalent to normalization in the Bayes rule for conditional probability distributions, and comes from decay of the total probability of the conditional trajectory \cite{ashida_2018}. For this reason the denominator is only zero when the trajectory occurs with probability zero, making this case unphysical. Notice that normalization also makes $\rho(t)$ invariant under constant shifts in $H$ both real and imaginary.

\subsubsection{Metric Formalism}
A Hamiltonian is said to be pseudo-Hermitian or $PT$-Symmetric if there exists Hermitian (generally non-unique) $\eta$ such that $H^\dag \eta = \eta H$. A satisfying $\eta$ is referred to as a metric, as it defines a modified inner product, $\langle \psi, \phi\rangle_\eta \equiv \langle\psi\vert\eta\vert\phi\rangle$, under which $H$ is Hermitian \cite{Bender_1998_real_spectra, Bender_2002_complex_extension, mostafazadeh_pseudo-hermiticity_2002}. Eigenvalues of pseudo-Hermitian $H$ come in complex conjugate pairs. If there further exists a positive definite $\eta$, then $H$ is said to be quasi-Hermitian or $PT$-unbroken and is guaranteed to have real eigenvalues. In this latter case we can decompose 
\begin{align}
\label{eq:SHS}
    H = S H_0 S^{-1}
\end{align}
 for Hermitian $S$ and $H_0$, where $H_0$ is isospectral to $H$. The similarity transform under $S = \eta^{-1/2}$ is sometimes referred to as the Dyson map \cite{Fring2016, Dyson1956}. In this case we can also decompose $U_t = S V_t S^{-1}$, for Unitary $V_t = e^{-iH_0t}$. Note that Hamiltonians resulting from Eq.~\ref{eq:H_eff} cannot be properly pseudo-Hermitian, but can be up to an overall imaginary shift, which generates equivalent normalized dynamics.

As an alternative to treating NH Hamiltonians as effective, one can view them as fundamental, acting in a modified Hilbert space $\mathscr{H}_\eta$ with inner product $\langle\cdot,\cdot\rangle_\eta$; this latter approach is referred to as the metric formalism \cite{mostafazadeh_pseudo-hermiticity_2002}. In this case $S$ acts as a map from the traditional Hilbert space $\mathscr{H}$ to $\mathscr{H}_\eta$, which is Unitary in the sense that $\langle S \psi, S \phi\rangle_\eta = \langle \psi \vert S \eta S \vert \phi \rangle = \langle \psi \vert \phi \rangle$. The two approaches to NH physics motivate two different extensions of CCs, which will both be considered in this paper.

\subsection{Non-Hermitian Transverse-Field Ising Model}
\label{sec:BG-TFIM}
Simulations in this paper use the 1D Imaginary Transverse-field Ising model (TFIM), a pseudo-Hermitian extension of the paradigmatic TFIM with well studied entanglement phases \cite{gopalakrishnan_entanglement_2021,Biella2021manybodyquantumzeno, barch2023, matsumoto_continuous_2020}. The Hamiltonian is:
\begin{align}
\label{eq:TFIM}
H = J\sum\limits_{j=1}^{L-1} \sigma_{j}^{z} \sigma_{j+1}^{z} + \sum\limits_{j=1}^{L} \left( g \sigma_{j}^{x} + h \sigma_{j}^{z}+i\gamma \sigma_j^y\right),
\end{align}
In the Hermitian ($\gamma=0$) limit, the model is integrable when $g$ or $h$ is $0$ and chaotic otherwise, in that it obeys random matrix spectral statistics and volume-law eigenstate entanglement \cite{PhysRevLett.106.050405,PhysRevLett.111.127205, wolfAreaLawsQuantum2008}. We use parameters $J=0.95$, $g=1$, and $h=0.5$ for numerical simulations, for which the model is chaotic in the Hermitian limit.

The NH TFIM can be generated by application of a Hermitian ($\gamma=0$) TFIM combined with continuous weak measurement of $y$ spins on all qubits, postselected to always yield $+1$. In this sense the non-Hermiticity parameter $\gamma$ quantifies measurement strength. When $\gamma < 1$ this model is quasi-Hermitian and can be decomposed according to Eq.~\ref{eq:SHS}. In this case $H_0$ is a Hermitian TFIM with $J_0=J,h_0=h$ and $g_0 = \sqrt{g^2-\gamma^2}$ \cite{matsumoto_continuous_2020}. $S$ can be written
\begin{align}
\label{eq:S-TFIM}
    S = \otimes_j e^{\frac{\beta}{2}\sigma_j^z}.
\end{align}
for $\beta = \text{atanh}(\frac{\gamma}{g})$. As we will see, the product nature of $S$ leads to similar entanglement properties between $H$ and $H_0$ as measured by modified CCs, though their properties differ under other measures such as operator entanglement \cite{barch2023}. Additionally, $\Vert S \Vert = e^{\frac{\beta n}{2}}$ grows exponentially in $n$, making placing tight bounds on similarity transformed quantities such as $\Vert U_t \Vert$ difficult.

\section{Connected Correlators on Non-Hermitian Systems}
\label{sec:CCs}

\begin{figure*}
    \begin{subfigure}{.245\textwidth}
        \includegraphics[width=0.9\linewidth]{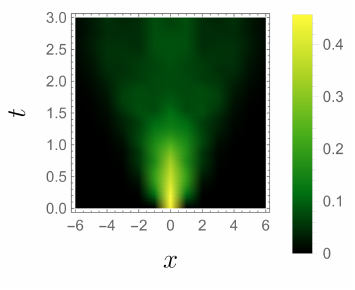}
        \caption{$\gamma=0$}
    \end{subfigure}%
    \begin{subfigure}{.245\textwidth}
        \includegraphics[width=0.9\linewidth]{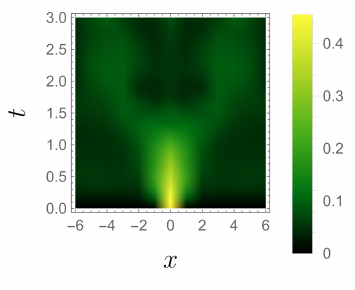}
        \caption{$\gamma=0.3$}
    \end{subfigure}%
    \begin{subfigure}{.245\textwidth}
        \includegraphics[width=0.9\linewidth]{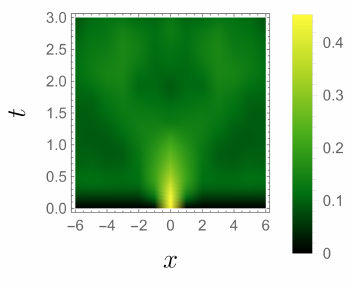}
        \caption{$\gamma=0.6$}
    \end{subfigure}%
    \begin{subfigure}{.245\textwidth}
        \includegraphics[width=0.9\linewidth]{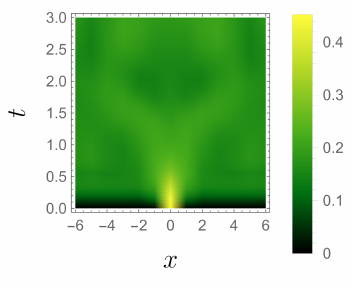}
        \caption{$\gamma=0.9$}
    \end{subfigure}%
    \caption{Plots of traditional $\vert\langle O_A(t),O_B(0)\rangle_c\vert$ for $A=0$ and $B=x$, averaged over $O_A,O_B \in \{\sigma^x, \sigma^y, \sigma^z\}$, for the NH TFIM (Eq.~\ref{eq:TFIM}), from an initial $\vert+^n\rangle$ state. As degree of non-Hermiticity $\gamma$ increases, the traditional CC ceases to demonstrate CC LR bound lightcones.}
    \label{fig:tradCC}
\end{figure*}

Many previously studied properties of CCs break down when extended to NH quantum systems, requiring a redefined CC to recover them. While the traditional equal-time CC can be written in terms of a state evolution, and so remains defined, both the unequal-time CC and previously mentioned CC LR bound require definition in terms of operator time evolution. This requires care, as the usual $O(t) = U_t^\dag O U_t$ leads to a breakdown of the automorphism property of time evolution, i.e. $(O_1 O_2)(t)\neq O_1(t) O_2(t)$, which necessary for interpretation of the equal-time CC in terms of a state evolution. It also leads to a breakdown of locality, as shown in Fig.~\ref{fig:tradCC}. In this section we first extend the traditional CC to a form which recovers some locality properties in NH systems, then propose an alternative CC derived from the metric formalism which recovers locality and even a CC LR bound in NH systems.

\subsection{Schr\"odinger CC}

As CCs are written in terms of correlators, we first extend the unequal-time disconnected correlator to NH systems. This can be done by interpreting operators as being applied at times $t',t$ to some state $\rho$ undergoing normalized time evolution under an NH Hamiltonian, where the operators are excluded from the normalizing factor \cite{Sergi_2015}. With this interpretation, the correlator is
\begin{align}
\begin{split}
    \langle O_1(t),O_2(t')\rangle_s &\equiv \frac{\langle U^\dag_{t} O_1 U_{t-t'} O_2 U_{t'}\rangle}{\langle U^\dag_{t} U_{t}\rangle}\\
    &= \frac{\langle O_1(t) \tilde O_2(t')\rangle}{\langle I(t)\rangle}
    \end{split}
\end{align}
where $\tilde O(t) = U_t^{-1} O U_t$ is a modified notion of operator time evolution in NH systems, for which $U_t^{-1}\neq U_t^\dag$. We refer to this as the Schr\"odinger correlator due to its motivation in terms of evolution of a state. This form has equivalence with the equal time correlator when $t=t'$, in which case the denominator is simply the trace normalization of $\rho(t)$ in Eq.~\ref{eq:mixed-evolution}. 

In extending this correlator to a CC, non-unitality of the $O(t)$ evolution requires additional caution. The standard definition equivalent to Eq.~\ref{eq:cc-herm} is no longer zero for product evolutions on product states, as one would like a CC to be. Instead, we define 
\begin{align}
\label{eq:CC-schro}
\begin{split}
    & \langle O_1(t),O_2(t')\rangle_{sc}\\
    &\equiv \langle O_1(t),O_2(t')\rangle_s - \langle O_1(t),I(t')\rangle_s \langle I(t),O_2(t')\rangle_s\\
    &=  \frac{\langle O_1(t) \tilde O_2(t')\rangle}{\langle I(t)\rangle} - \frac{\langle O_1(t) \rangle\langle I(t) \tilde O_2(t')\rangle}{\langle I(t)\rangle^2}
    \end{split}
\end{align}
which recovers the desired locality behavior by explicitly subtracting out the product case. Despite this, the $O(t)$ evolution violates LR bounds, which means this CC can not be shown to obey the CC LR bound in general. In fact in the case $t'=0$ it is equivalent to the traditional CC (Eq.~\ref{eq:cc-herm}), for which the lightcone breakdown is plotted in Fig.~\ref{fig:tradCC}.

\subsection{Metric CC}
An alternative CC, which does recover LR bounds, can be derived from the metric formalism of $PT$-Symmetric systems \cite{mostafazadeh_2003, mostafazadeh_pseudo-hermiticity_2002}. For pseudo-Hermitian $H$ and Hermitian $\eta$ such that $\eta H = H^\dag \eta$, consider the modified Hilbert space $\mathscr{H}_\eta$ with inner product $\langle \psi, \phi\rangle_\eta \equiv \langle\psi\vert\eta\vert\phi\rangle$. Within $\mathscr{H}_\eta$ the \textit{bra} dual to \textit{ket} $\vert\psi\rangle$ is $\langle\psi\vert\eta$, so when $\rho\in\mathcal{L}(\mathscr{H})$ represents an ensemble of kets in $\mathscr{H}$, the density matrix in $\mathcal{L}(\mathscr{H}_\eta)$ describing the same ensemble of kets in $\mathscr{H}_\eta$ is \cite{karuvade_2022}
\begin{align}
\label{eq:sigma}
\sigma \equiv \frac{\rho\eta}{\langle\eta\rangle}
\end{align}
where $\langle\eta\rangle = \text{Tr}[\rho\eta]$ gives trace normalization in $\mathcal{L}(\mathscr{H}_\eta)$. 

When $\eta$ is invertible the modified adjoint in $\mathcal{L}(\mathscr{H}_\eta)$, $O^\# \equiv \eta^{-1} O^\dag \eta$, gives the canonical "Unitary" operator time evolution $U_t^\# O U_t = U_t^{-1} O U_t = \tilde O (t)$ for pseudo-Unitary $U = e^{-iHt}$. With this evolution and state $\sigma$, the standard unequal-time correlator in $\mathscr{H}_\eta$ is then
\begin{align}
\label{eq:corr-metric-sigma}
    \langle O_1(t),O_2(t')\rangle_\eta = \langle \tilde O_1(t) \tilde O_2(t')\rangle_\sigma
    = \frac{\langle \eta \tilde O_1(t) \tilde O_2(t')\rangle}{\langle \eta\rangle}
\end{align}
and the associated CC is
\begin{align}
\begin{split}
\label{eq:cc-metric}
    \langle O_1(t),O_2(t')\rangle_{\eta c} &\equiv \\
    \frac{\langle\eta \tilde O_1(t)\tilde O_2(t')\rangle}{\langle\eta\rangle} &- \frac{\langle\eta \tilde O_1(t)\rangle \langle\eta \tilde O_2(t')\rangle}{\langle\eta\rangle^2}
\end{split}
\end{align}

We will refer to this as the Metric CC, and henceforth interpret it as a modified CC within the standard Hilbert space $\mathscr{H}$. Notice that since $U \rho\eta U^{-1} = U\rho U^\dag \eta$, the equal-time Metric CC can be written in terms of time-evolved state $\rho(t)$. This would not be the case if we were to use the $\tilde O(t)$ evolution without $\eta$, or to take $\sigma = S\rho S^{-1}$ for $S = \eta^{-1/2}$, even though the latter can be interpreted as representing the same state as $\rho$ \cite{karuvade_2022}. 

Note we are only guaranteed the existence of $\eta > 0$, and thus non-degenerate $\langle\cdot,\cdot\rangle_\eta$, when $H$ is quasi-Hermitian. For indefinite $\eta$ this correlator is defined but less well motivated, and can diverge as $\langle\eta\rangle\rightarrow 0$. 

An extension of both CCs to the $n$-partite case is discussed in Appendix~\ref{app:n-part} via a generalization of the CC generating function \cite{Tran_2017, sylvester1975}. Interestingly, the $n$-partite Metric CC takes a more natural form than the Schr\"odinger CC, stemming from the former's role as standard CC in $\mathscr{H}_\eta$.

\section{Connection to Entanglement}
\label{sec:CC-entang}
In this section we extend the relation between CCs and entanglement mentioned in Sec.~\ref{sec:BG-CC-entang} by decomposing $\delta\rho \equiv \rho - \rho_A\otimes\rho_B$ as a sum of equal-time CCs between operators on subsystems $A$ and $B$. This allows extension of the CC LR bound to functions of $\delta\rho$, in particular $\Vert\delta\rho\Vert_2$ and mutual information $I(A;B)$. Additionally, we show how this decomposition extends to the Metric CC, where we find it holds for $\delta\sigma$ in general but $\delta\rho$ only when $A,B$ bipartition the entire system. Because of this difference, we note that the Metric CC LR bound does not extend to one on $\delta\rho$, as the bound requires some distance between $A$ and $B$ for information to propagate.

Let $\rho$ denote the state of the $AB$ (sub)system, which will be a reduced state when $A,B$ do not form a system bipartition. Consider operator Schmidt decomposition \cite{zanardi_entanglement_2001-1,lupo2008bipartite,aniello2009relation} of $\rho$,
\begin{align}
    \rho = \sum_{i=1}^r \lambda_i\ \Gamma_A^i\otimes\Gamma_B^i
\end{align}
where $r$ is the Schmidt rank of $\rho$, $\lambda_i = \langle \Gamma_A^{i\dag}\otimes\Gamma_B^{i\dag}\rangle \geq 0$, and $\{\Gamma_X^i\}_i$ are an orthonormal set $\text{Tr}[\Gamma_X^{i\dag} \Gamma_X^j]=\delta_{ij}$ which can be extended to an operator basis on $X=A,B$. Let $C_i \equiv \langle \Gamma_A^{i\dag}, \Gamma_B^{i\dag}\rangle_c$ be the traditional equal-time CC in Eq.~\ref{eq:cc-herm}, which can be equivalently taken with respect to the full system state or reduced state $\rho$.  We may write
\begin{align}
\begin{split}
\label{eq:delrho-decomp}
    \rho &= \sum_{i=1}^r \left\langle \Gamma_A^{i\dag}\otimes\Gamma_B^{i\dag} \right\rangle  \Gamma_A^i\otimes\Gamma_B^i\\
    &= \sum_{i=1}^r \Big(\langle \Gamma_A^{i\dag}\rangle\langle\Gamma_B^{i\dag}\rangle + C_i\Big)\ \Gamma_A^i\otimes\Gamma_B^i\\
    &= \rho_A\otimes\rho_B + \delta\rho
    \end{split}
\end{align}

with $\delta\rho \equiv \rho - \rho_A\otimes\rho_B = \sum_{i=1}^r C_i\ \Gamma_A^i\otimes\Gamma_B^i$. Then we find, for example,
\begin{align}
\label{eq:delrho-norm-bound}
\begin{split}
    \Vert\delta\rho\Vert_2^2 &= \sum_{i,j=1}^r \overline C_i C_j \text{Tr}\left[\Gamma_A^{\dag i}\Gamma_A^j\otimes\Gamma_B^{\dag i}\Gamma_B^j\right] = \sum_{i=1}^r \vert C_i\vert^2
\end{split}
\end{align}

For time evolved $\rho(t)$, we can use Eq.~\ref{eq:CC-LR-bound} to bound $\vert C_i\vert\leq \bar c\ e^{-\frac{L-2vt}{\chi'}}$, which holds $\forall t$ regardless of time dependence implicit in $\Gamma_A^i,\ \Gamma_B^i$. Then by bounding $r \leq d_\text{min}^2$ for time-independent $d_\text{min} \equiv \text{min}(d_A, d_B)$, we find

\begin{align}
\label{eq:delrho-LR-bound}
    \Vert\delta\rho(t)\Vert_2 \leq \bar c\ d_\text{min} e^{-\frac{L-2vt}{\chi'}} 
\end{align}

The result is an operator-free LR bound on depending only on $\rho(t)$.

\subsection{Mutual Information Bound}
As shown in Appendix~\ref{app:info-bound}, when $\rho$ is full matrix rank (not to be confused with Schmidt rank), the mutual information between $A,B$ can be bound in terms of $\delta\rho$ as
\begin{align}
    I(A;B) \leq \Vert \text{log}(k\rho)\Vert_2\Vert\delta\rho\Vert_2
\end{align}
where $k>0$ is a positive free parameter, which can be minimized to get the tightest bound. The logarithmic term can be shown to be constant, so when a CC LR bound is present, this allows placement of an LR bound on mutual information at time t:
\begin{align}
\begin{split}
\label{eq:info-LR-bound}
    I(A;B)(t) &\leq \Vert \text{log}(k\rho_0)\Vert_2\Vert\delta\rho(t)\Vert_2\\
    &\leq \tilde c\ d_\text{min}e^{-\frac{L - 2vt}{\chi'}}
    \end{split}
\end{align}

where $\tilde c = \overline c \Vert\text{log}(k\rho_0)\Vert_2$ for initial state $\rho_0$. The full rank requirement on $\rho_0$ ensures that $\Vert\text{log}(k\rho_0)\Vert_2$ is bounded. This condition is satisfied for generic $\rho_0$, for example by thermal subsystem states which can emerge even in a closed overall system \cite{popescu_entanglement_2006, D_Alessio_2016, borgonovi_quantum_2016, dymarsky2018}. 

\begin{figure*}
    \begin{subfigure}{.245\textwidth}
        \includegraphics[width=0.9\linewidth]{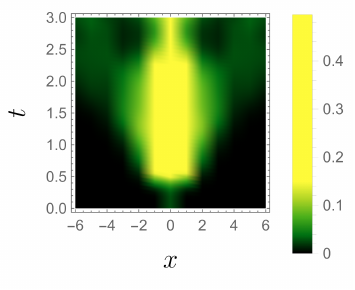}
        \caption{$\gamma=0$}
    \end{subfigure}%
    \begin{subfigure}{.245\textwidth}
        \includegraphics[width=0.9\linewidth]{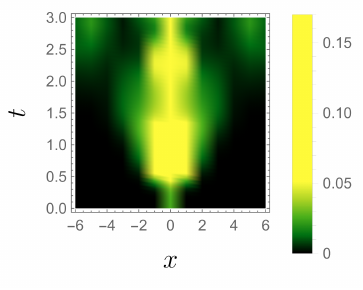}
        \caption{$\gamma=0.3$}
    \end{subfigure}%
    \begin{subfigure}{.245\textwidth}
        \includegraphics[width=0.9\linewidth]{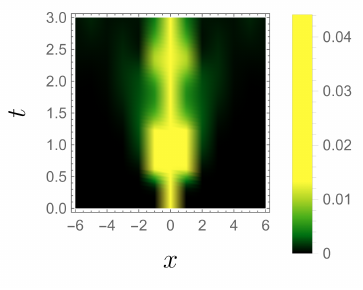}
        \caption{$\gamma=0.6$}
    \end{subfigure}%
    \begin{subfigure}{.245\textwidth}
        \includegraphics[width=0.9\linewidth]{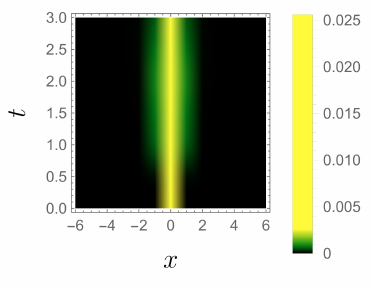}
        \caption{$\gamma=0.9$}
    \end{subfigure}%
    \caption{Plots of $I(A;B)$ for $A=0$ and $B=x$ for the NH TFIM (Eq.~\ref{eq:TFIM}), from an initial Gibbs state at inverse temperature $\beta=3$ under Hamiltonian $-S_x$. $I(A;A)=H(A)$ at $A=x=0$ is included for completeness. One sees mutual information obeys an LR bound for all $\gamma$. However, as non-Hermiticity increases, the mutual information sharply decreases, especially for $\vert x\vert>1$.}
    \label{fig:info}
\end{figure*}

As an example, consider thermal state $\rho_0 = e^{-\beta H'_{AB}}/Z$ under Hamiltonian $H'_{AB}$, with $Z = \text{Tr}[e^{-\beta H'_{AB}}]$. We can pick $k=Z$ so that $k\rho_0 = e^{-\beta H'_{AB}}$, and
\begin{align}
        \tilde c = \overline c \Vert \text{log}\left(e^{-\beta H'_{AB}}\right)\Vert_2 = \overline c \beta \Vert H'_{AB}\Vert_2
\end{align}
Notice that $\Vert H'_{AB}\Vert_2$ grows as $O(d_{AB})$ rather than $O(d)$, making it bounded for fixed $\vert A \vert, \vert B \vert$ even in the thermodynamic limit. Such a case is illustrated in Fig.~\ref{fig:info}, where mutual information growth obeys an LR bound following a quench from an initial thermal state of $H'_{AB}=-\sigma_A^x-\sigma_B^x$.

\subsection{Entanglement in Non-Hermitian Systems}

The aforementioned $\delta\rho$ decomposition is agnostic to time evolution, and so generalizes to NH systems. However, as the CC LR bound breaks down, Eq.~\ref{eq:delrho-LR-bound} no longer holds in general. This motivates asking whether the Metric CC, which we will show does obey LR bounds, can be used instead to expand $\delta\rho$.

The result, as shown in Appendix~\ref{app:entang-metric}, is that when $A, B$ form a system bipartition and $\eta=\eta_A\otimes\eta_B$, $\delta\rho$ can be expanded as a linear combination of at most $4r$ Metric CCs. However when $A,B$ do not form a system bipartition, the presence of $\eta$ in the Metric CC prevents it from being written as a function of reduced state cleanly. That is, for tripartite system $ABC$, let $\rho$ now represent the overall state and $\rho_{AB}$ the state on $AB$. Even with $\eta=\eta_{AB}\otimes\eta_C$, we find

\begin{align}
\begin{split}
    \langle\eta O_A O_B \rangle_\rho &= \text{Tr}[O_A O_B \text{Tr}_C[\rho \eta]]\\
    &= \text{Tr}[\eta_{AB} O_A O_B \text{Tr}_C[\rho \eta_C]]\\
    &\neq \langle\eta_{AB} O_A O_B \rangle_{\rho_{AB}}
\end{split}
\end{align}

The same issue arises for CCs. Intuitively, $\eta_C$ modifies the entanglement properties of $\rho$ between the $A,B$ subsystems. Thus we cannot prove an LR bound on $\Vert\delta\rho(t)\Vert_2$ or mutual information in NH systems in the same way as for Hermitian ones, even though we still see one numerically in Fig.~\ref{fig:info}. This distinction has an interesting application, discussed in Sec.~\ref{sec:app-entang}.

We can alternatively consider bounding $\delta\sigma(t)$. From Eqs.~\ref{eq:corr-metric-sigma}, \ref{eq:cc-metric} we see that the Metric CC on state $\rho(t)$ is equivalent to a traditional CC on modified state $\sigma(t)$. This lets us use the decomposition in Eq.~\ref{eq:delrho-decomp} to write $\delta\sigma(t)$ as a sum of Metric CCs, and so extend the Metric CC LR bound to a bound on $\Vert\delta\sigma(t)\Vert_2$. While $\sigma(t)$ has interpretation as a state in $\mathscr{H}_\eta$, its physical significance is unclear, so for now this bound is primarily of mathematical interest.

\section{LR Bounds of Modified CCs}
\label{sec:LR}
We will now show the main result: that the Metric CC obeys an LR bound even in local $PT$-Symmetric systems. We also find that the unequal-time Schr\"odinger CC obeys an LR bound when its initial state is a thermal state of the evolution Hamiltonian. 
Before that, notice that as any quasi-Hermitian $H$ will be Hermitian in modified Hilbert space $\mathscr{H}_\eta$, the generated dynamics can be shown to obey an LR bound in $\mathscr{H}_\eta$ in terms of modified operator norm $\Vert\cdot\Vert_\eta$. However the practical meaning of $\Vert\cdot\Vert_\eta$ is unclear, and as shown in Appendix.~\ref{app:UinvLR}, conversion to the traditional operator norm picks up overhead potentially exponential in $n$. This makes extending the LR bound this way of questionable utility. Instead, in this section we will show the LR bound on the Metric and Schr\"odinger CCs has no such overhead.
\subsection{Metric CC}
When $H$ is quasi-Hermitian under product $\eta$ and sufficiently local, an LR bound holds for the Metric CC. This can be shown by decomposing $H = S H_0 S^{-1}$ for Hermitian $H_0$ and $S = \eta^{-1/2}$. Then $U_t = S V_t S^{-1}$ for $V_t = e^{-iH_0 t}$, and one can rewrite the Metric correlator in terms of Unitary evolutions on modified state and operators:

\begin{align}
\label{eq:cc-metric-decomp}
    \langle O_A(t),O_B(t')\rangle_{\eta} = \langle V_{t}^\dag \hat O_A V_{t-t'} \hat O_B V_{t'} \rangle_{\hat \sigma} 
\end{align}

where $\hat O_X \equiv S^{-1} O_X S$ and $\hat\sigma \equiv S^{-1} \sigma S = \langle\eta\rangle_\rho^{-1}S^{-1}\rho S^{-1}$. This can be interpreted as using the map $S: \mathscr{H}\rightarrow\mathscr{H}_\eta$ to rewrite evolution of operators on $\mathscr{H}_\eta$ in terms of evolution of operators on $\mathscr{H}$. For product $S$ decomposable as $S = S_X\otimes S_{\overline X}$ the similarity transform of operators is locality-preserving: $\hat O_X = S^{-1}_X O_X S_X$. In this case the entire Metric CC may be written as a traditional CC with Unitary evolution $\hat O (t) \equiv V^\dag_t \hat O V_t$:

\begin{align}
\label{eq:metric-CC-dyson}
\begin{split}
    &\langle O_A(t),O_B(t')\rangle_{\eta c}\\
    &= \langle \hat O_A(t) \hat O_B(t')\rangle_{\hat\sigma} - \langle \hat O_A(t) \rangle_\sigma \langle \hat O_B(t')\rangle_{\hat \sigma}
    \end{split}
\end{align}

Thus, we can directly apply locality results on traditional CCs with Unitary evolution, such as LR bounds and exponential clustering \cite{Tran_2017,bravyi_lieb-robinson_2006,Nachtergaele_2006}. Using the bound in Eq.~\ref{eq:CC-uneq-LR-bound}, we find

\begin{align}
\label{eq:CC-metric-LR}
    \vert \langle \hat O_A(t),\hat O_B(t')\rangle_{\eta c} \vert\leq \bar c\ \Vert \hat O_A\Vert \Vert \hat O_B \Vert e^{-\frac{L-v(t+t')}{\chi'}}
\end{align}

with $\Vert \hat O_X \Vert \leq \Vert S_X\Vert \Vert S^{-1}_X \Vert$ for product $S$, taking $\Vert O_X\Vert\leq 1$ without loss of generality. $\Vert S_X\Vert,\Vert S_X^{-1}\Vert$ can grow as $O(d_X)$, but will be finite for fixed $\vert X\vert$ even in the thermodynamic limit.

This bound requires that $H_0$ be composed of sufficiently local terms as in the original LR bound, and that $\hat\sigma$ exhibit finite correlation length. The former follows from locality of $H$, as when $S$ is a tensor product over subsystems of fixed locality, such a single qubits, the locality properties of $H$ and $H_0$ will be equivalent. That is, let $H = \sum_R h_R$ and take $S = \otimes_{Q} S_Q$ for $R,Q$ of fixed locality. With $R' \equiv \cup_{Q\cap R\neq\varnothing} Q'$, we can write $S = S_{R'}\otimes S_{\overline {R'}}$ for each $R'$. Then 
\begin{align}
    H_0 = S^{-1} (\sum_R h_R) S = \sum_R S^{-1}_{R'} h_R S_{R'} 
\end{align}
which satisfies the conditions for an LR bound due to fixed locality of $R'$. For example when $Q$ are single qubits, $R'=R$ simply.

Finite correlation length of $\hat\sigma$ is similarly motivated, though difficult to prove in general due to lack of cancellation. In the case that $\rho$ is the ground state of gapped $H$, $\hat \sigma$ is the ground state of the gapped isospectral $H_0$, guaranteeing finite correlation length \cite{Nachtergaele_2006}. 

\begin{figure*}
    \begin{subfigure}{.245\textwidth}
        \includegraphics[width=0.9\linewidth]{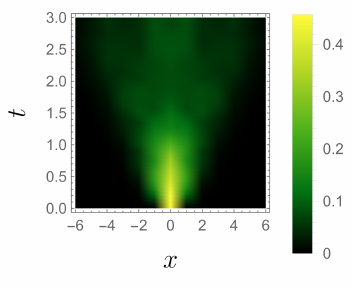}
        \caption{$\gamma=0$}
    \end{subfigure}%
    \begin{subfigure}{.245\textwidth}
        \includegraphics[width=0.9\linewidth]{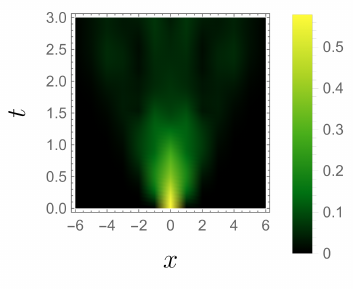}
        \caption{$\gamma=0.3$}
    \end{subfigure}%
    \begin{subfigure}{.245\textwidth}
        \includegraphics[width=0.9\linewidth]{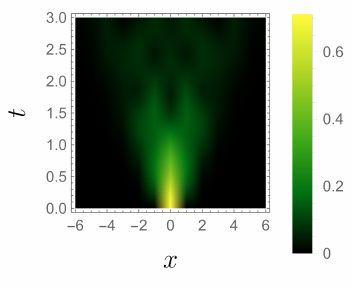}
        \caption{$\gamma=0.6$}
    \end{subfigure}%
    \begin{subfigure}{.245\textwidth}
        \includegraphics[width=0.9\linewidth]{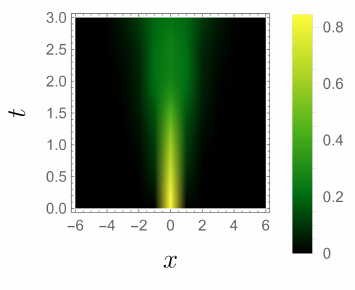}
        \caption{$\gamma=0.9$}
    \end{subfigure}%
    \caption{Plots of $\langle O_A(t),O_B(0)\rangle_{\eta c}$ for $A=0$ and $B=x$, summed over $O_A,O_B \in \{\sigma^x, \sigma^y, \sigma^z\}$. The lightcone structure is preserved as $\gamma$ increases, unlike for the Schr\"odinger CC. Narrowing of the lightcone can be attributed to a decreased LR velocity, as for $\gamma\approx1$ the Hamiltonian transitions from volume-law to area-law entanglement generation \cite{barch2023}}
    \label{fig:unequal-metric}
\end{figure*}

An example quasi-Hermitian system with product Metric is the NH TFIM \cite{matsumoto_continuous_2020,barch2023}. For this system we can pick $S$ as in Eq.~\ref{eq:S-TFIM}, for which $\Vert S_X \Vert = \Vert S^{-1}_X\Vert = e^{\vert X\vert/2}$. This choice results in the lightcone structure in Fig.~\ref{fig:unequal-metric}, indicative of the Metric CC LR bound.

\subsection{Schr\"odinger CC}
The Schr\"odinger correlator can be decomposed similarly to Eq.~\ref{eq:cc-metric-decomp}, yielding
\begin{align}
\label{eq:cc-decomp-schro}
    \langle O_A(t),O_B(t')\rangle_{s} = \frac{\langle\eta\rangle\langle V_{t}^\dag (S O_A S) V_{t-t'} \hat O_B V_{t'} \rangle_{\hat \sigma}}{\langle I(t)\rangle}
\end{align}

Unlike $\hat O_A$, $S O_A S$ can be generally nonlocal, preventing placement of an LR bound on the Schr\"odinger CC in general. The exception is that when $\rho$ is a thermal state of $H$, the Schr\"odinger correlator is time invariant like in the Hermitian case: 

\begin{align}
\label{eq:thermal-equiv}
   \langle O_A(t) \tilde O_B(t')\rangle_{s} = \langle O_A(0) \tilde O_B(t-t')\rangle_{s}
\end{align}

In this case the time dependence of the Schr\"odinger CC is limited to $\tilde O_B(t-t')$, and the CC can be decomposed in terms of $\hat O_A$, $\hat O_B$ as

\begin{align}
\label{eq:LR-schro-thermal}
\begin{split}
    \langle O_A(t),O_B(t')\rangle_{sc} &= \langle O_A(0), O_B(t-t')\rangle_{s c}\\
    &= \langle \hat O_A, \hat O_B(t'-t)\rangle_{c, \hat \rho}
\end{split}
\end{align}

For $\hat\rho = S^{-1}\rho S$. Then if one assumes that $\hat \rho$ has finite correlation length, the CC LR bound for Unitary evolution applies:

\begin{align}
    \left\vert \langle O_A(t),O_B(t')\rangle_{sc}\right\vert \leq \bar c\ \Vert \hat O_A\Vert \Vert \hat O_B \Vert e^{-\frac{L-v\vert t'-t\vert}{\chi'}}
\end{align}

Unfortunately this bound is only nontrivial for the unequal time CC, limiting its application towards describing properties of the state such as entanglement. Finite correlation length of $\hat\rho$ is also harder to motivate than for $\hat\sigma$, as $\hat\rho$ is not a proper state.

\subsection{Generalization}
Quasi-Hermiticity of $H$ under product $\eta$ is not strictly necessary for existence of an LR bound on the Metric CC; a bound on $\Vert\tilde O_X(t)\Vert = \Vert U_t^{-1}O_X U_t\Vert$ is sufficient. Given existence of such a bound, one can retrace the steps in Sec.~\ref{sec:BG-CC} to derive an LR bound on both the Metric CC and the thermal state Schr\"odinger CC, without relying on a decomposition of $U_t$. The presence of such an LR bound is well motivated, since the breakdown of LR bounds for $O(t)$ in Sec.~\ref{sec:BG-LR} does not occur for the $\tilde O(t)$ evolution:

\begin{align*}
\label{eq:LR-Uinv-motivation}
    \dot{\tilde{O}}_A\big\vert_{t=0} &= i [H_{\mathrm{eff}}, \tilde O_A] = i\sum_{R\cap A \neq \varnothing} [H_R + i\Gamma_R, \tilde O_A]
\end{align*}

However while we see evidence for such an LR bound in Fig.~\ref{fig:LR_Uinv_normalized}, a general analytic proof is not evident, and even in the quasi-Hermitian case of Appendix~\ref{app:UinvLR} the bound incurs overhead potentially exponential in system size.

\section{Applications}
\label{sec:applications}
Here we discuss two more practical results: how the previous Metric CC LR bound can place a necessary condition of the set of local quasi-Hermitian Hamiltonians capable of generating long-range entangled states, and how traditional and metric CCs may be measured as a series of POVMs.
\subsection{Preparing Long-range Entangled States}
\label{sec:app-entang}
Limits on entanglement growth under local Hamiltonians has motivated interest in  non-local entanglement protocols such as gate teleportation and entanglement swapping, which augment Unitary evolution with measurement and classical communication \cite{nielsen_chuang_2010}. NH Hamiltonians provide an alternative method of generating non-local entanglement, where classical communication is replaced by postselection. The Metric CC LR bound restricts this long range entangling power, but by finding cases where the bound is trivially satisfied, we can derive a necessary condition on the class of quasi-Hermitian Hamiltonians capable of generating long range entangled states in short times.

Consider first Hermitian systems, where the CC LR bound (Eq.~\ref{eq:CC-LR-bound}) can be solved for $t$ to lower bound the time needed to generate a given state. For example, the $n$-qubit GHZ state has $\langle\sigma_1^z,\sigma_n^z\rangle_c = 1$, while the CC LR bound states that $\vert\langle \sigma^z_1(t),\sigma^z_n(t)\rangle_c\vert\leq\overline c\ e^{-\frac{n-2vt}{\chi'}}$. Thus as $n\rightarrow\infty$, the time needed to generate the GHZ state an initial state with finite correlation length becomes infinite.

In quasi-Hermitian systems, this logic can be extended to the Metric CC. Solving the Metric CC LR bound in Eq.~\ref{eq:CC-metric-LR} for $t=t'$ yields
\begin{align}
    t \geq \frac{\chi'}{2v}\text{log}\left(\frac{\vert \langle \hat O_A(t),\hat O_B(t')\rangle_{\eta c} \vert}{\bar c\ \Vert \hat O_A\Vert \Vert \hat O_B \Vert}\right) +\frac{L}{2v}
\end{align}
In particular as $L\rightarrow\infty$, $t\rightarrow\infty$. The exception is that when the Metric CC is zero for all $t$, the LR bound is trivially satisfied and no longer restricts $t$. As we now show, this can occur even for entangled $\rho$.


As mentioned at the end of Sec.~\ref{sec:CC-entang}, the Metric CC quantifies the entanglement of $\sigma$, rather than $\rho$ itself. For a tripartite system $ABC$ the two can be different even for product $\eta$, and if $\sigma_{AB} = \text{Tr}_C[\sigma] = \langle\eta\rangle^{-1}\text{Tr}_C[\rho\eta]$ is product between $A$ and $B$, the Metric CC will be zero for all operator pairs $O_A, O_B$. In particular if $\sigma(t)_{AB} = \langle\eta\rangle^{-1} \text{Tr}_C[\rho(t)\eta]$ is product for all $t$, the Metric CC LR bound will become trivial.

For given $\rho$ and $H$, and $t\in\mathbb{R}$, consider the two sets
\begin{align}
    \begin{split}
        \mathbb{F}_{\rho(t)} &= \{\eta\ :\ \sigma(t)_{AB}\ \text{product on }A\vert B\}\\
        \mathbb{G}_H &= \{\eta\ :\ H^\dag\eta=\eta H^\dag, \eta\text{ product on }A\vert B\vert C\}
    \end{split}
\end{align}
For $H,\rho_0=\rho(0)$ with $\mathbb{G}_H\subseteq\cap_t \mathbb{F}_{\rho(t)}$, the Metric CC LR bound is trivial for all $t$ and no longer restricts the entangling power of $H$. Thus if one wishes to generate long range entanglement under quasi-Hermitian Hamiltonians, it is this set of $H,\rho_0$ that must be considered. An example system where $\sigma(t)_{AB}$ stays product while $H$ entangles $\rho(t)$ is provided in Appendix~\ref{app:app-entang}.

\subsection{Measurement as POVM}
\label{sec:measure}
NH time evolutions on states can be generated by combining Unitary evolution with weak measurement and postselection \cite{ashida_non-hermitian_2020}, or by changing system metric by applying $S$ as a positive operator valued measurement (POVM) \cite{karuvade_2022}. However, unequal-time correlators introduce ambiguity from the interaction between applied operators and state normalization, such as occurs in Refs.~\cite{Sergi_2015, barch2023}, and in operator time evolution itself.  Though the CCs described in this paper are written in terms of operator time evolution, here we show how they can be measured as the joint success probability of a series of POVMs and state evolutions, where the state evolutions need not be normalized, resolving ambiguity. An alternative approach is seen in Ref.~\cite{matsumoto_embedding_2022}, where the Metric correlator on a thermal state is mapped to a standard correlator on an extended Hermitian system via projective measurements.

Application of operator $O_i$ with $\Vert O_i\Vert\leq 1$ via POVM to an initial pure state $\vert\psi\rangle$ succeeds with probability $P(O_i) = \Vert O_i\vert\psi\rangle\Vert^2$. Similarly, evolving $\vert\psi\rangle$ under well-behaved $U_t$ generated by an NH Hamiltonian, e.g. as generated by weak measurement conditioned on certain outcomes, succeeds with probability $P(U_t) = \Vert U_t\vert\psi\rangle\Vert^2$. To see this, one could imagine a Trotter expansion of $U_t$ alternating Unitary evolution with near-identity POVMs. By well-behaved $U_t$, we mean that success probability must be non-increasing over time, which follows from assuming success at each time step is independent. This then imposes a condition on NH Hamiltonian $H$. Let $H = H_0-i\Gamma$ for Hermitian $H_0,\Gamma$, then
\begin{align}
    \begin{split}
        \frac{d}{dt}\Vert \vert\psi(t)\rangle\Vert^2 &= i\langle\psi(t)\vert H^\dag-H\vert\psi(t)\rangle\\
        &= -2\langle\psi(t)\vert\Gamma\vert\psi(t)\rangle
    \end{split}
\end{align}
so success probability $\Vert\vert\psi(t)\rangle\Vert^2$ is nonincreasing regardless of $\vert\psi\rangle$ when $\Gamma\geq0$. This occurs naturally for $H$ coming from no-jump Lindblad evolution (Eq.~\ref{eq:H_eff}) and can be equivalent to pseudo-Hermitian evolution up to a constant shift in $\Gamma$. Interestingly even when the minimum eigenvalue of $\Gamma$ is zero, $\Vert U_t \Vert$ and thus $\Vert\vert\psi(t)\rangle\Vert^2$ may still exhibit exponential decay owing to $H_0$ mixing eigenspaces of $\Gamma$. 

Using Bayes rule, the probability of applying a pair of operators via separate POVMs (or non-Unitary evolutions) can be written cleanly as

\begin{align}
    \begin{split}
        P(O_A, O_B) &= P(O_A\vert O_B)\cdot P(O_B)\\
        &= \left\Vert O_A \frac{O_B\vert\psi\rangle}{\Vert O_B\vert\psi\rangle\Vert}\right\Vert^2\cdot\Vert O_B\vert\psi\rangle\Vert^2\\
        &= \Vert O_A O_B\vert\psi\rangle\Vert^2\\
        &= P(O_A O_B)
    \end{split}
\end{align}

Treating measurement of $\Pi_\psi = \vert\psi\rangle\langle\psi\vert$ similarly, we find, e.g., 
\begin{align}
    \begin{split}
        P(\Pi_\psi, U_t^\dag, O_A, U_t, O_B) &= \Vert \Pi_\psi U_t^\dag O_A U_t O_B \vert\psi\rangle\Vert^2\\
        &= \vert \langle\psi\vert U_t^\dag O_A U_t O_B\vert\psi\rangle \vert^2
    \end{split}
\end{align}

This extends similarly to all other expectation values used in this paper, and all CCs can be calculated in terms of success probabilities, each of which can be measured independantly. Decay of success probability with time and number of operations can make the process take exponentially many measurements to accurately estimate expectation values, but by accounting for this we have escaped the hidden costs inherent in the usual postselection. In the case of quasi-Hermitian evolution, $S$ and $S^{-1}$ can be applied independently as in Ref.~\cite{karuvade_2022}, or absorbed into existing operators applied as POVMs, leaving all necessary time evolution Unitary.

\section{Discussion}
\label{sec:conclusion}

While non-Hermitian quantum systems have a long history, their study under the lens of quantum information is relatively new, leaving much room for generalization of existing methods and bounds. Interaction between the metric formalism of non-Hermitian systems and information theoretic quantities is especially interesting, as the formalism provides a unified way to study a large class of non-Hermitian systems of interest. Here we used this approach to extend the well-studied LR bound to connected correlators on local $PT$-Symmetric systems. This is significant as the usual LR bound was previously shown to break down in these systems, and even when applied to the unital $\tilde O(t)$ evolution picks up exponential overhead not present in the connected correlator LR bound.

The connected correlator LR bound here discussed promises insight into many questions of entanglement in non-Hermitian systems, such as the recently popular topic of measurement-induced phase transitions, a phenomena describable by non-Hermitian Hamiltonians  \cite{PhysRevB.98.205136,PhysRevX.9.031009,PhysRevB.99.224307,PhysRevB.100.134306,PhysRevX.10.041020,PhysRevLett.125.030505,PhysRevX.11.011030, agarwal2023recognizing}, or the topic of non-Hermitian quantum chaos \cite{Efetov1997,Chalker1997,Fyodorov1997,fyodorov1998universality,Fyodorov2003, barch2023}. As seen in Fig.~\ref{fig:unequal-metric}, the Metric connected correlator reveals a sharp drop in LR velocity for the imaginary Transverse-Field Ising Model near critical measurement strength $\gamma \approx 1$, seemingly detecting the Hamiltonian's phase transition from chaotic to integrable that occurs at this point \cite{gopalakrishnan_entanglement_2021, barch2023}. Furthermore the Metric connected correlator diverges at exceptional points, where $\eta$ becomes indefinite, allowing it to detect measurement induced criticality.

The LR bounds shown on entanglement and mutual information in Sec.~\ref{sec:CC-entang} are novel in that they give operator-free bounds on information scrambling. In addition, they have implications towards the construction of efficient MPS representation of $\rho$ in Hermitian systems, and thus towards simulability and quantum computational complexity of these systems \cite{preskill_2012}. The extension to bounding entanglement of $\sigma$ in non-Hermitian systems means these systems, traditionally believed more difficult to simulate due to LR bound violation, may be simulable via MPS representation of $\sigma$ rather than $\rho$. Utility of using the entanglement of $\sigma$ to bound that of $\rho$ remains a topic for further study, as even in the bipartite case the bound has overhead proportional to $\Vert\eta\Vert$, which can grow exponentially. 

Finally, it is worth noting that any evolution with measurement and classical action can be written as an ensemble of postselected quantum evolutions, potentially allowing results from non-Hermitian systems to generalize to this unconditional yet non-Markovian case. In particular, while Sec.~\ref{sec:app-entang} places restrictions on long range entanglement generation under non-Hermitian evolution, it may be possible to extend this to bound generation of long range entanglement via classical communication, e.g. as used in entanglement swapping and teleportation. 

\section{Acknowledgements}
This research was partially supported by the ARO MURI grant W911NF-22-S-0007. B.B. would like to thank Namit Anand, Daniel Lidar, and Paolo Zanardi for insightful discussions during the writing of this paper.

\appendix
\begin{appendix}
\numberwithin{equation}{section}

\section{LR bound on unequal-time CCs}
\label{app:CCLR-uneq}
This section extends the proof of LR bounds for equal-time CCs in Eq.~\ref{eq:CC-LR-bound} and Ref.~\cite{bravyi_lieb-robinson_2006} to the unequal-time case. Given

\begin{align}
\begin{split}
    &\left\vert \langle O_A(t),O_B(t')\rangle_c \right\vert\\
    &\leq \left\vert \langle O_A^l(t), O_B^l(t')\rangle_c\right\vert + c\vert A\vert e^{\frac{l-vt}{\xi}}+c\vert B\vert e^{\frac{l'-vt'}{\xi}}\\
    &\leq \tilde c\ e^{-\frac{L-l-l'}{\chi}} + c\vert A\vert e^{\frac{l-vt}{\xi}}+c\vert B\vert e^{\frac{l'-vt'}{\xi}}
    \end{split}
\end{align}

we can pick the optimal $l = (\xi L+\chi v t + \xi v(t-t'))/\chi'$ and $l' = (\xi L+\chi v t' + \xi v(t'-t))/\chi'$. Plugging this in yields Eq.~\ref{eq:CC-uneq-LR-bound}. This bound can be extended to the $n$-partite $n$-time case following the steps of Ref.~\cite{Tran_2017}, but the interplay of multiple distances and times makes the results quite cumbersome.

\section{$n$-partite extension of CCs}
\label{app:n-part}

The CCs in Sec.~\ref{sec:CCs} extend to the $n$-partite case via a generalization of the generating function in Eq.~\ref{eq:BG-gen-func} and Refs.~\cite{sylvester1975,Tran_2017} to the unequal-time case. Recall that the standard CC can be extended to $n$ operators by writing it as a sum over partitions $P$ of the set $\{1,...,n\}$ \cite{sylvester1975,Tran_2017} as

\begin{align}
\label{eq:cc_n}
    \langle O_1,...,O_n\rangle_c = \sum_{P} g(|P|) \prod_{p\in P} \left\langle \prod_{i\in p} O_i\right\rangle
\end{align}
for $g(x+1)=(-1)^x x!$. This can further be written in terms of a generating function as

\begin{align}
\label{eq:BG-gen-func}
    \langle O_1,...,O_n\rangle_c \equiv\frac{\partial^n}{\partial \lambda_1...\partial \lambda_n}\text{ln}\left\langle e^{\sum_i \lambda_i O_i}\right\rangle \bigg\vert_{\vec \lambda = 0}
\end{align}

For Hermitian systems this can simply be written in terms of time evolved operators as, e.g., $O_i \rightarrow O_i(t_i)$. However in NH systems the Heisenberg picture becomes ambiguous, so here we reconstruct the generating function in terms of evolution of states.

\subsection{General form of CC generating function}
\label{app:generating-func}

Consider $n$ copies of the system with some $O_i$ acting on the $i^{th}$ copy. Define the cyclic permutation over system copies $P = \prod_{i=2}^{n} \mathbb{S}_{i-1,i}$ for SWAP operator $\mathbb{S}$, and extended state $\rho' = P (\rho \otimes I^{\otimes n-1})$ (akin to as written in Ref.~\cite{anand_brotocs_2022}). 

For time-evolution channel $\mathcal{E}_{\vec t} = \otimes_i \mathcal{E}_{t_i}$, and $\langle \vec O(\vec t)\rangle_c \equiv \langle O_1(t_1),...,O_n(t_n)\rangle_c$, a general generating function and $n$-partite CC are given by

\begin{align}
\label{eq:general-generator}
    \begin{split}
    \langle \vec O(\vec t)\rangle_c &= \frac{\partial^n}{\partial \lambda_1...\partial \lambda_n}\text{ln}\Big\langle \otimes_i e^{\lambda_i O_i} \Big\rangle_{\mathcal{E}_{\vec t}(\rho')} \bigg\vert_{\vec \lambda = 0}\\
    &= \frac{\partial^n}{\partial \lambda_1...\partial \lambda_n} \text{ln}\left\langle \otimes_i \mathcal{E}^\dag_{t_i}\left( e^{\lambda_i O_i}\right) \right\rangle_{\rho'} \bigg\vert_{\vec \lambda = 0}\\
    &= \sum_P \frac{g\big(|P|\big)}{\big\langle \prod_{i} \mathcal{E}^\dag_{t_i}(I) \big\rangle^{|P|}} \prod_{p\in P} \Big\langle \prod_{i} \mathcal{E}^\dag_{t_i}(O_i^{p_i})\Big\rangle
    \end{split}
\end{align}
for indicator variables $p_i\in\{0,1\}$ so that $O_i^{p_i}=O_i$ iff $i\in p$ and $O_i^{p_i}=I$ otherwise. In the case of Unitary (or any trace-preserving) $\mathcal{E}$, the adjoint channel $\mathcal{E}^\dag$ is unital and this reduces to the usual form in Eq.~\ref{eq:cc_n} with $O_i\rightarrow O_i(t_i)$. Eq.~\ref{eq:general-generator} is used to derive the generating functions for the Schr\"odinger and Metric CCs (Eq.~\ref{eq:generator-schro} and Eq.~\ref{eq:generator-metric}). For the Shcr\"odinger CC we take $\mathcal{E}_{t_1}(\cdot) = U_{t_1}\cdot U_{t_1}^\dag = U_{t_1}\cdot\eta U_{t_1}^{-1}\eta^{-1}$ and $\mathcal{E}_{t_i}(\cdot) = U_{t_i}\cdot U_{t_i}^{-1}$ for $i\neq 1$, while for the Metric CC we use the $\mathcal{E}_{t_i} = U_{t_1}\cdot U_{t_i}^{-1}\ \forall i$, with $\rho\rightarrow\sigma=\frac{\rho\eta}{\langle\eta\rangle}$.

The proof of vanishing CCs for product states in Ref.~\cite{Tran_2017} holds for this general form as well in the case of product time evolution. The $n$-partite CC derived here is also inherently normalized even when $\rho$ is not, making trace normalization of $\rho'(\vec t\ )$ optional when extending to non-trace-preserving time evolution.

\subsection{$n$-partite Schr\"odinger and Metric CCs}

We can use this general form to construct generating functions for the $n$-partite Schr\"odinger and Metric CCs, yielding forms guaranteed to be zero for product evolutions on product states. These generating functions can be written in terms of a modified initial state $\sigma' = P (\rho\eta \otimes I^{\otimes n-1})$ and time evolution $\tilde \sigma'(\vec t) = U_{\vec t} \sigma' U_{\vec t}^{-1}$. With this, we get the generating functions

\begin{align}
     \langle \vec O (\vec t)\rangle_{sc} &= \frac{\partial^n}{\partial \lambda_1...\partial \lambda_n}\text{ln}\Big\langle \eta^{-1}_1\left(\otimes_i e^{\lambda_i O_i}\right) \Big\rangle_{\tilde \sigma'(\vec t)} \bigg\vert_{\vec \lambda = 0}
\end{align}
and
\begin{align}
    \langle \vec O (\vec t)\rangle_{\eta c} &= \frac{\partial^n}{\partial \lambda_1...\partial \lambda_n}\text{ln}\Big\langle \otimes_i e^{\lambda_i O_i} \Big\rangle_{\tilde \sigma'(\vec t)} \bigg\vert_{\vec \lambda = 0}
\end{align}

for the $n$-partite Schr\"odinger and Metric CCs, respectively, where $\eta^{-1}_1 \equiv \eta^{-1}\otimes I^{\otimes n-1}$. Writing the equal-time $n$-partite Metric CC in this form allows extension of the LR bound, shown for the two-partite Metric CC in Sec.~\ref{sec:LR}, to the $n$-partite case via the results of Ref.~\cite{Tran_2017}.

From the generating functions we derive that the explicit form of the $n$-partite Schr\"odinger CC is

\begin{align}
\label{eq:generator-schro}
    \langle \vec O(\vec t)\rangle_{\eta c} = \sum_P \frac{g\big(|P|\big)}{\langle I(t_1) \rangle^{|P|}} \prod_{p\in P} \Big\langle O_1^{p_1}(t_1) \prod_{i\in p,i\neq 1} \tilde O_i(t_i)\Big\rangle
\end{align}

for indicator variable $p_1$ such that $O_1^{p_1}=O_1$ iff $1\in p$ and $O_1^{p_1}=I$ otherwise. For the Metric CC it is

\begin{align}
\label{eq:generator-metric}
    \langle \vec O(\vec t)\rangle_{\eta c} = \sum_P \frac{g\big(|P|\big)}{\langle\eta\rangle^{|P|}} \prod_{p\in P} \Big\langle \eta \prod_{i\in p} \tilde O_i(t_i)\Big\rangle
\end{align}

\section{Entanglement, Information, and CCs}
\label{app:entang-bound}

In this section we show that when $\rho$ is full rank, mutual information can be upper bounded in terms of $\Vert\delta\rho\Vert_2$ for $\delta\rho=\rho-\rho_A\otimes\rho_B$, and when applicable in terms of an LR bound. Additionally, we discuss two extensions of the decomposition of $\delta\rho$ in Sec.~\ref{sec:CC-entang}, the first to an arbitrary operator basis and the second to $PT$-Symmetric systems.

\subsection{Bounding Mutual Information}
\label{app:info-bound}

Here we show that mutual information $I(A;B)$ can be upper bounded in terms of $\Vert\delta\rho\Vert_2$. Using the CC bound on $\Vert\delta\rho\Vert_2$ in Eq.~\ref{eq:delrho-norm-bound}, one can then bound it in terms of a sum of CCs, and when dynamics are sufficiently local in terms of an LR bound. The requirement is that the reduced state $\rho$ on the two subsystems $A,B$ is full rank so that $\text{log}(\rho)$ is bounded from below. This condition is generically satisfied, for example by thermal states which can emerge in subsystems of even a closed overall system \cite{popescu_entanglement_2006, D_Alessio_2016, borgonovi_quantum_2016, dymarsky2018}.

Let $H(X) = -\text{Tr}[\rho_X\text{log}(\rho_X)]$ denote entropy on the $X$ subsystem. Mutual information is defined as 
\begin{align}
    & I(A;B) = H(A)+H(B)-H(AB)\\ \nonumber
    &= \text{Tr}[\rho\text{log}(\rho)] - \text{Tr}[\rho_A\otimes\rho_B\text{log}(\rho_A\otimes\rho_B)]\\ \nonumber
    &= \text{Tr}[\rho_A\otimes\rho_B(\text{log}(\rho)-\text{log}(\rho_A\otimes\rho_B))]+\text{Tr}[\delta\rho\text{log}(\rho)]\\ \nonumber
    &= \text{Tr}[\delta\rho\text{log}(\rho)] - D(\rho_A\otimes\rho_B \Vert \rho)
\end{align}

where in the second line we use $\text{Tr}[\rho_A\otimes\rho_B\text{log}(\rho_A\otimes\rho_B)] = \text{Tr}[\rho_A\text{log}(\rho_A)]+\text{Tr}[\rho_B\text{log}(\rho_B)] = -H(A)-H(B)$. The second term in the final line is $\leq 0$ owing to non-negativity of relative entropy $D$. For the first term note that since $\text{Tr}[\delta\rho]=0$ we can shift $\text{log}(\rho) \rightarrow \text{log}(k\rho) = \text{log}(\rho)+\text{log}(k)\mathbb{I}$ for $k>0$ without changing the overall quantity. Then using H\"older's inequality,
\begin{align}
    \begin{split}
        I(A;B) &\leq \text{Tr}[\delta\rho\text{log}(k\rho)]\\
        &\leq \Vert \delta\rho\Vert_2 \Vert \text{log}(k\rho) \Vert_2
    \end{split}
\end{align}

Interestingly, as $\text{log}(U\rho U^\dag) = U\text{log}(\rho)U^\dag$ and the 2-norm is invariant under unitary rotations, $\Vert\text{log}(k\rho)\Vert_2$ is time-independent under unitary evolution and may be written in terms of initial state $\rho_0$. Thus from the LR bound on $\Vert\delta\rho(t)\Vert_2$ in Eq.~\ref{eq:delrho-LR-bound}, we find an LR bound on mutual information
\begin{align}
\label{eq:info-LR-bound-app}
\begin{split}
    I(A;B)(t) &\leq \Vert\delta\rho(t)\Vert_2\Vert\text{log}(k\rho_0)\Vert_2\\
    &\leq d_\text{min} \overline c\ \Vert\text{log}(k\rho_0)\Vert_2\ e^{-\frac{L - 2vt}{\chi'}}
    \end{split}
\end{align}

\subsection{Alternative choice of basis}

Given that the Schmidt decomposition can be hard to calculate, the bound in Sec.~\ref{sec:CC-entang} can also be found using an arbitrary orthonormal product basis for $AB$, $\{\Gamma_A^i, \Gamma_B^j\}$. Let $C_{ij} = \langle \Gamma_A^{i\dag}, \Gamma_B^{j\dag}\rangle_c$ be the traditional equal-time CC. Then
\begin{align}
\begin{split}
    \rho &= \sum_{ij} \left\langle \Gamma_A^{i\dag}\otimes\Gamma_B^{j\dag} \right\rangle  \Gamma_A^i\otimes\Gamma_B^j\\
    &= \sum_{ij} \Big(\langle \Gamma_A^{i\dag}\rangle\langle\Gamma_B^{j\dag}\rangle + C_{ij}\Big)\Gamma_A^i\otimes\Gamma_B^j\\
    &= \rho_A\otimes\rho_B + \delta\rho
    \end{split}
\end{align}
for $\delta\rho = \sum_{ij} C_{ij}\ \Gamma_A^i\otimes\Gamma_B^j$.

\subsection{Metric CC}
\label{app:entang-metric}

Recall that the Metric CC on $\rho$ is equivalent to traditional CC on modified state $\sigma$ (Eq.~\ref{eq:corr-metric-sigma}), or on modified state $\hat\sigma$ and operators $\hat O_X$ (Eq.~\ref{eq:metric-CC-dyson}). This allows decomposition of $\delta\sigma$ and $\delta\hat\sigma$ in terms of a sum of Metric CCs on $\rho$. While $\delta\sigma$ and $\delta\hat\sigma$ have value in the metric formalism their physical meaning is unclear, and it is difficult to write $\delta\rho$ in terms of $\delta\sigma$ or $\delta\hat\sigma$ in general.

Luckily, in the case that $A,B$ bipartition the entire system we can decompose the traditional CC as a linear combination of Metric CCs, and thus write $\delta\rho$ as a sum of Metric CCs directly.

To see this, take $\eta = \eta_A\otimes\eta_B$, let $\langle\eta\rangle=1$ without loss of generality, and drop tensor product notation for compactness. Then we can decompose
\begin{align}
    \begin{split}
        &\langle O_A O_B\rangle = \langle\eta(\eta_A^{-1}O_A\eta_B^{-1}O_B)\rangle\\
        &= \langle\eta_A^{-1}O_A,\eta_B^{-1}O_B\rangle_{\eta c} + \langle O_A\eta_B\rangle\langle\eta_A O_B\rangle
    \end{split}
\end{align}
Plugging this in for each expectation value in the traditional CC and then performing a lengthy calculation involving the equality $1 - \langle\eta_A\rangle\langle\eta_B\rangle = \langle\eta_A^{-1},\eta_B^{-1}\rangle_{\eta c}$ yields
\begin{align}
        &\langle O_A,O_B\rangle_c = \langle O_A O_B\rangle - \langle O_A I_B\rangle \langle I_A O_B\rangle\\
        \nonumber
        &= \langle\eta_A^{-1}O_A,\eta_B^{-1}O_B\rangle_{\eta c} + \langle O_A\eta_B\rangle\langle\eta_A O_B\rangle \langle\eta_A^{-1},\eta_B^{-1}\rangle_{\eta c}\\
        &- \langle\eta_A^{-1}O_A,\eta_B^{-1}\rangle_{\eta c}\langle O_B\rangle - \langle O_A\eta_B\rangle\langle\eta_A\rangle\langle\eta_A^{-1},\eta_B^{-1}O_B\rangle_{\eta c} \nonumber
\end{align}
The original decomposition in Eq.~\ref{eq:delrho-decomp} writes $\delta\rho$ as a sum of $r$ CCs, where $r$ is the operator Schmidt rank of $\rho$, so $\delta\rho$ can be written as a linear combination of at most $4r$ different Metric CCs. 

\section{LR bound on $[\tilde O_A(t), O_B]$}
\label{app:UinvLR}

As shown in Eq.~\ref{eq:LR-Uinv-motivation}, the LR bound breakdown mentioned in Sec.~\ref{sec:BG-LR} does not occur for the $\tilde O(t)$ evolution used in the Schr\"odinger and Metric CCs. This motivates study of whether an LR bound can be established on commutators of the form $[\tilde O_A(t),O_B]$ in general. While we see such a bound numerically in the TFIM (Figs.~\ref{fig:LR_Uinv_normalized}, \ref{fig:LR_Uinv}), a general proof picks up exponential overhead even in the quasi-Hermitian case, making it unwieldy.

\subsection{Analytics}
In the case that $H$ is quasi-Hermitian, we can decompose $U_t = S V_t S^{-1}$ for Unitary $V_t$ and let $\hat O = S^{-1} O S$. Then we find

\begin{align}
\begin{split}
\label{eq:LR-commu-Uinv}
    \Vert [ \tilde O_A (t),& O_B] \Vert = \Vert [S V_t^\dag \hat O_A V_t S^{-1}, O_B] \Vert\\
    &= \Vert S [\hat O_A(t), \hat O_B] S^{-1} \Vert\\
    &\leq \Vert S \Vert \Vert S^{-1} \Vert \Vert \hat O_A\Vert \Vert \hat O_B\Vert c N_\text{min}e^{-\frac{L-vt}{\xi}}
    \end{split}
\end{align}

So while the LR bound holds, the utility of it depends on $\Vert S \Vert \Vert S^{-1} \Vert$, which may be exponentially extrinsically large, e.g. for the imaginary TFIM with S given in Sec.~\ref{sec:BG-TFIM}, where $\Vert S \Vert = \Vert S^{-1} \Vert = e^{\frac{\beta n}{2}}$. This highlights that the Metric CC is a special case where nonlocal parts of $S$ cancel out.

\subsection{Numerics}
\begin{figure*}
    \begin{subfigure}{.245\textwidth}
        \includegraphics[width=0.9\linewidth]{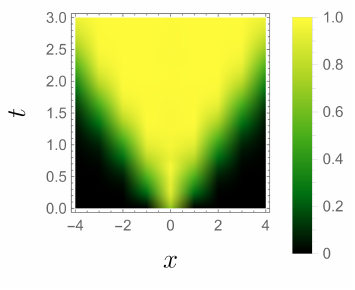}
        \caption{$\gamma=0$}
    \end{subfigure}%
    \begin{subfigure}{.245\textwidth}
        \includegraphics[width=0.9\linewidth]{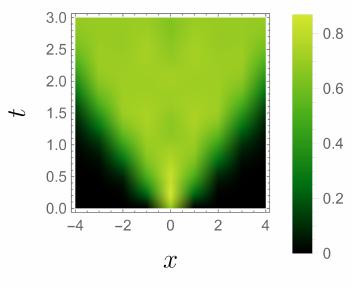}
        \caption{$\gamma=0.3$}
    \end{subfigure}%
    \begin{subfigure}{.245\textwidth}
        \includegraphics[width=0.9\linewidth]{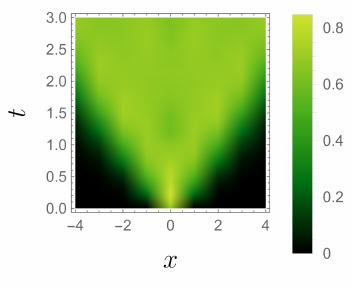}
        \caption{$\gamma=0.6$}
    \end{subfigure}%
    \begin{subfigure}{.245\textwidth}
        \includegraphics[width=0.9\linewidth]{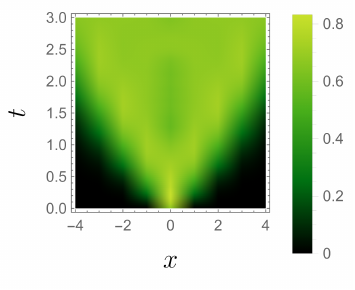}
        \caption{$\gamma=0.9$}
    \end{subfigure}%
    \caption{Plots of $\frac{1}{2} \Vert [ \tilde O_A(t), O_B]\Vert / \Vert \tilde O_A(t) \Vert$ for $A=1$ and $B=x$, where $\tilde O_A(t) = U^{-1} O_A U$ is evolved under the imaginary TFIM. Results are averaged over $O_A,O_B \in \{\sigma^x, \sigma^y, \sigma^z\}$. Plots demonstrate that this operator evolution preserves the LR bound.}
    \label{fig:LR_Uinv_normalized}
\end{figure*}

\begin{figure*}
    \begin{subfigure}{.245\textwidth}
        \includegraphics[width=0.9\linewidth]{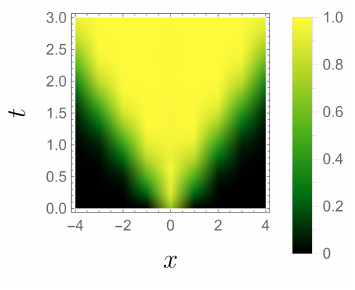}
        \caption{$\gamma=0$}
    \end{subfigure}%
    \begin{subfigure}{.245\textwidth}
        \includegraphics[width=0.9\linewidth]{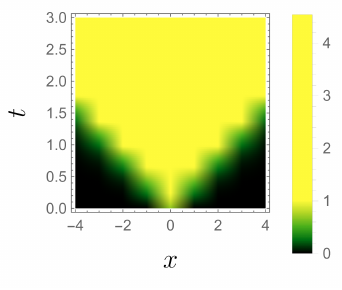}
        \caption{$\gamma=0.3$}
    \end{subfigure}%
    \begin{subfigure}{.245\textwidth}
        \includegraphics[width=0.9\linewidth]{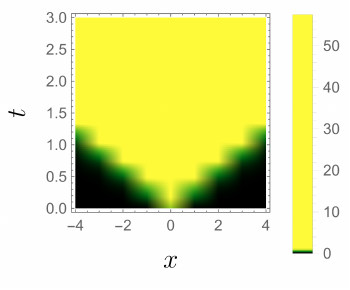}
        \caption{$\gamma=0.6$}
    \end{subfigure}%
    \begin{subfigure}{.245\textwidth}
        \includegraphics[width=0.9\linewidth]{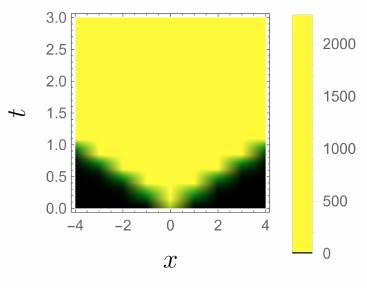}
        \caption{$\gamma=0.9$}
    \end{subfigure}%
    \caption{Plots of $\frac{1}{2} \Vert [ \tilde O_A(t), O_B]\Vert$ for $A=1$ and $B=x$, where $\tilde O_A(t) = U^{-1} O_A U$ is evolved under the imaginary TFIM. Results are averaged over $O_A,O_B \in \{\sigma^x, \sigma^y, \sigma^z\}$. Plots demonstrate that growth of $\Vert \tilde O_A(t)\Vert$ is dominated by exponential decay arising from the LR bound, and lightcone dynamics dominate, albeit at increased effective $v$.}
    \label{fig:LR_Uinv}
\end{figure*}

Though a tight analytic LR bound is elusive, we do numerically see lightcone structure in $\Vert [ \tilde O_A(t), O_B]\Vert / \Vert \tilde O_A(t) \Vert$ in Fig.~\ref{fig:LR_Uinv_normalized}. The caveat is that $\Vert \tilde O(t) \Vert$ may grow in time. For quasi-Hermitian $H$ it saturates, but may saturate to a value in $O(2^n)$. However in certain cases the exponential decay of the LR bound dominates for large enough distances, and so in Fig.~\ref{fig:LR_Uinv} we see an LR lightcone for even non-normalized $\Vert[\tilde O_A(t),O_B]\Vert$, but with increased effective LR velocity.

\section{Entanglement Preparation Example}
\label{app:app-entang}
As an example case where the Metric CC LR bound is trivially satisfied, as discussed in Sec.~\ref{sec:app-entang}, consider tripartite system $ABC$. Begin with an operator $O_C > 0$ not proportional to the identity, and use Gram-Schmidt to extend the set $\{O_C, \mathbb{I_C}\}$ to an orthonormal basis $\{\Gamma_C^i\}_{i=1}^{d^2}$ for $\mathscr{L(H}_C)$. Notice that since $\mathbb{I}_C\in\text{span}\{\Gamma_C^1,\Gamma_C^2\}$, $\text{Tr}[\Gamma_C^k] = \text{Tr}[\mathbb{I}_C\Gamma_C^k ] = 0$ for $k>2$. Then we can decompose arbitrary $\rho$ as
\begin{align}
\label{eq:rho-basis-decomp-examp}
    \rho = \sum_{k=1}^{d^2} a_k \rho_{AB}^k\otimes\Gamma_C^k
\end{align}
The $\rho_{AB}^k$ need not be density matrices in general, but for now assume they are. Taking $\eta_C = \Gamma_C^1$, basis orthonormality gives us
\begin{align}
    \begin{split}
        \text{Tr}_C[\rho] &= a_1\text{Tr}[\Gamma_C^1]\rho_{AB}^1 + a_2\text{Tr}[\Gamma_C^2]\rho_{AB}^2\\
        \text{Tr}_C[\rho\eta_C] &= a_1\rho_{AB}^1
    \end{split}
\end{align}
Any metric CC of the form $\langle O_A,O_B\rangle_{\eta c}$ depends only on $\text{Tr}_C[\rho\eta]$. Thus if we take $\rho_{AB}^1$ product and $\eta=\eta_A\otimes\eta_B\otimes\eta_C$ for the given $\eta_C$, such a Metric CC will always be zero, trivially satisfying the Metric CC LR bound. Notice also this places no restriction on $\rho_{AB}^2$, which can be entangled to make $\rho_{AB}=\text{Tr}_C[\rho]$ entangled.

As an example of generation of long range entanglement, consider $\rho$ as in Eq.~\ref{eq:rho-basis-decomp-examp} with $a_1=1$ and $a_2=0$, and product $\rho_{AB}^1$. In this case $\text{Tr}_C[\rho]$ is product. Pick some $H = H_C$ satisfying $H_C^\dag\eta - \eta H_C = 0$ for $\eta = \eta_C$. Finally, define $\delta\Gamma_C^k = H_C\Gamma_C^k-\Gamma_C^k H_C^\dag$ and let $C_{jk} = \text{Tr}[\Gamma_C^{j\dag}\delta\Gamma_C^k]$ be coefficients of $\delta\Gamma_C^k$ in the $\{\Gamma_C^j\}$ basis. Note that for all $k$, Hermiticity of $\Gamma_C^1 \sim O_C > 0$ gives 
\begin{align}
        C_{1k} = C_{k1} = \text{Tr}[\Gamma_C^{k\dag}\big(H_C\Gamma_C^1-\Gamma_C^1 H_C^\dag\big)] = 0
\end{align}

Then, excluding normalization,
\begin{align}
\begin{split}
        \text{Tr}_C[\partial_t \rho] &= -i\text{Tr}_C[H_C\rho - \rho H_C^\dag] = -i\sum_{k\neq 2} a_k\rho_{AB}^k\text{Tr}[\delta\Gamma_C^k]\\
        &= -i \sum_{k\neq2} a_k\rho_{AB}^k\left(\sum_{j} C_{jk}\text{Tr}\left[\Gamma_C^j\right]\right)\\
        &= -i \sum_{k>2} a_k\rho_{AB}^k\left(C_{2k}\text{Tr}\left[\Gamma_C^2\right]\right)\\
\end{split}
\end{align}
using $\text{Tr}[\Gamma_C^j]=0$ for $j>2$. Thus for entangled $\rho_{AB}^k$, $H_C$ generates entanglement between the $A,B$ subsystems at short times. 

\end{appendix}

\bibliographystyle{apsrev4-1}
\bibliography{bib2, bib}

\end{document}